\documentclass[aps,prl,reprint,superscriptaddress, nofootinbib]{revtex4-2}

\usepackage{graphicx}
\usepackage{dcolumn}
\usepackage{bm}
\usepackage{hyperref}
\usepackage{amsmath}
\usepackage{amssymb}
\usepackage{listings}
\usepackage{xcolor}
\usepackage{url}
\usepackage{footmisc}
\begin{document}

\preprint{APS/123-QED}

\title{CUORE Data Release for ML Applications: Pulse Shape Analysis Dataset}
\newcommand{\cuoreUSC}{\affiliation{Department of Physics and Astronomy, University of South Carolina, Columbia, SC 29208, USA}}
\newcommand{\cuoreVT}{\affiliation{Center for Neutrino Physics, Virginia Polytechnic Institute and State University, Blacksburg, Virginia 24061, USA}}
\newcommand{\cuoreLegnaro}{\affiliation{INFN -- Laboratori Nazionali di Legnaro, Legnaro (Padova) I-35020, Italy}}
\newcommand{\cuoreBolognaINFN}{\affiliation{INFN -- Sezione di Bologna, Bologna I-40127, Italy}}
\newcommand{\cuoreSapienza}{\affiliation{Dipartimento di Fisica, Sapienza Università di Roma, Roma I-00185, Italy}}
\newcommand{\cuoreRomaINFN}{\affiliation{INFN -- Sezione di Roma, Roma I-00185, Italy}}
\newcommand{\cuoreGSSI}{\affiliation{Gran Sasso Science Institute, L'Aquila I-67100, Italy}}
\newcommand{\cuoreLNGS}{\affiliation{INFN -- Laboratori Nazionali del Gran Sasso, Assergi (L'Aquila) I-67100, Italy}}
\newcommand{\cuoreBerkeley}{\affiliation{Department of Physics, University of California, Berkeley, CA 94720, USA}}
\newcommand{\cuoreMiBINFN}{\affiliation{INFN -- Sezione di Milano Bicocca, Milano I-20126, Italy}}
\newcommand{\cuoreMiBU}{\affiliation{Dipartimento di Fisica ``Giuseppe Occhialini'', Universit\`{a} degli Studi di Milano-Bicocca, Milano I-20126, Italy}}
\newcommand{\cuoreGenovaINFN}{\affiliation{INFN -- Sezione di Genova, Genova I-16146, Italy}}
\newcommand{\cuoreGenovaU}{\affiliation{Dipartimento di Fisica, Università di Genova, Genova I-16146, Italy}}
\newcommand{\cuoreFudan}{\affiliation{Key Laboratory of Nuclear Physics and Ion-beam Application (MOE), Institute of Modern Physics, Fudan University, Shanghai 200433, China}}
\newcommand{\cuoreLBNLnuc}{\affiliation{Nuclear Science Division, Lawrence Berkeley National Laboratory, Berkeley, CA 94720, USA}}
\newcommand{\cuorePavia}{\affiliation{INFN -- Sezione di Pavia, Pavia I-27100, Italy}}
\newcommand{\cuoreBolognaU}{\affiliation{Dipartimento di Fisica e Astronomia, Alma Mater Studiorum -- Università di Bologna, Bologna I-40127, Italy}}
\newcommand{\cuoreIJCLab}{\affiliation{Université Paris-Saclay, CNRS/IN2P3, IJCLab, 91405 Orsay, France}}
\newcommand{\cuoreFrascati}{\affiliation{INFN -- Laboratori Nazionali di Frascati, Frascati (Roma) I-00044, Italy}}
\newcommand{\cuoreCalPoly}{\affiliation{Physics Department, California Polytechnic State University, San Luis Obispo, CA 93407, USA}}
\newcommand{\cuoreSJTU}{\affiliation{INPAC and School of Physics and Astronomy, Shanghai Jiao Tong University; Shanghai Laboratory for Particle Physics and Cosmology, Shanghai 200240, China}}
\newcommand{\cuoreYale}{\affiliation{Wright Laboratory, Department of Physics, Yale University, New Haven, CT 06520, USA}}
\newcommand{\cuoreUCLA}{\affiliation{Department of Physics and Astronomy, University of California, Los Angeles, CA 90095, USA}}
\newcommand{\cuoreJHU}{\affiliation{Department of Physics and Astronomy, The Johns Hopkins University, 3400 North Charles Street Baltimore, MD, 21211}}
\newcommand{\cuoreMIT}{\affiliation{Massachusetts Institute of Technology, Cambridge, MA 02139, USA}}
\newcommand{\cuoreIRFU}{\affiliation{IRFU, CEA, Université Paris-Saclay, F-91191 Gif-sur-Yvette, France}}
\newcommand{\cuoreBerkeleyNE}{\affiliation{Department of Nuclear Engineering, University of California, Berkeley, CA 94720, USA}}
\newcommand{\cuoreCassino}{\affiliation{Dipartimento di Ingegneria Civile e Meccanica, Università degli Studi di Cassino e del Lazio Meridionale, Cassino I-03043, Italy}}
\newcommand{\cuorePitt}{\affiliation{Department of Physics and Astronomy, University of Pittsburgh, Pittsburgh, PA 15260, USA}}
\newcommand{\cuorePadova}{\affiliation{INFN -- Sezione di Padova, Padova I-35131, Italy}}
\newcommand{\cuoreLBNLeng}{\affiliation{Engineering Division, Lawrence Berkeley National Laboratory, Berkeley, CA 94720, USA}}

\author{D.~Q.~Adams}                \cuoreUSC
\author{C.~Alduino}                 \cuoreUSC
\author{K.~Alfonso}                 \cuoreVT
\author{F.~T.~Avignone III}         \cuoreUSC
\author{O.~Azzolini}                \cuoreLegnaro
\author{G.~Bari}                    \cuoreBolognaINFN
\author{F.~Bellini}                 \cuoreSapienza\cuoreRomaINFN
\author{G.~Benato}                  \cuoreGSSI\cuoreLNGS
\author{M.~Beretta}                 \cuoreBerkeley
\author{M.~Biassoni}                \cuoreMiBINFN
% ...



\collaboration{CUORE Collaboration}

\date{\today}

\begin{abstract}
We present a public dataset from the CUORE (Cryogenic Underground Observatory for Rare Events) experiment, designed to support the development and benchmarking of Artificial Intelligence and Machine Learning (AI/ML) algorithms for cryogenic calorimeter data analysis. CUORE uses TeO$_2$ cryogenic calorimeters to measure particle energy depositions as thermal fluctuations. This dataset contains thermal pulses measured during calibration data taking. Each data point is provided as a one-dimensional, time-series array corresponding to a thermal pulse, accompanied by a binary classification label to distinguish between single-pulse events and pile-up events with two or more pulses; pulse normalization parameters and relevant metadata are also included, with all data stored in HDF5 format. This data release enables the testing of supervised learning approaches to pulse shape analysis, pile-up identification, and related tasks in the context of rare-event searches with cryogenic calorimeters.
\end{abstract}

\maketitle
\section{Introduction}

Neutrinoless double-beta (0$\nu\beta\beta$) decay is a hypothetical lepton-number-violating process whose discovery would confirm the Majorana nature of the neutrino and provide a mechanism for explaining the matter-antimatter asymmetry of the universe~\cite{Majorana1937,Sakharov1967}. The CUORE (Cryogenic Underground Observatory for Rare Events) experiment searches for 0$\nu\beta\beta$ decay of $^{130}$Te using an array of 988 TeO$_2$ crystal calorimeters operated at millikelvin temperatures at the Laboratori Nazionali del Gran Sasso (LNGS) in Italy~\cite{CUORE2022Nature}.

CUORE has accumulated over 2~tonne$\cdot$yr of TeO$_2$ exposure, setting a lower limit on the $^{130}$Te 0$\nu\beta\beta$ decay half-life of $T_{1/2} > 3.5 \times 10^{25}$~yr (90\% C.I.)~\cite{CUORE2tyr}.
The data release presented here is an excellent resource for the AI/ML community as it combines the high-quality, time-series data from CUORE with well-understood, physics-based labels.

In response to the growing demand for open scientific datasets~\cite{NAIRR2023} and the development of new analysis techniques, the CUORE Collaboration is releasing a curated subset of its data. This release specifically targets the problem of \emph{pulse classification}: distinguishing clean single pulses from pile-up pulses, where two or more pulses are present in the same detector event window. 

\section{The CUORE Detector and Pulse Formation}

\subsection{Detector Overview}

The CUORE detector consists of 988 $5\times5\times5$~cm$^3$ TeO$_2$ crystal calorimeters. The crystals are arranged in 19 towers of 52 crystals each; the active detector mass is 742~kg. The CUORE detector is hosted inside a custom dilution refrigerator, which cools the crystals to a base temperature of $\sim$10~mK. At these temperatures, the heat capacity of the crystals follow the Debye law $C(T) \propto T^3$, making the temperature fluctuations from energy depositions measurable.

The crystals serve as both $\beta\beta$ source (containing $^{130}$Te at $\sim$34\% natural isotopic abundance) and as the calorimetric absorbers~\cite{fiorini}. An NTD (Neutron Transmutation Doped) germanium thermistor~\cite{Haller1984} is glued to each crystal to measure the temperature change resulting from energy depositions. 

\subsection{Pulse Shape}

When a particle deposits energy $E$ in a TeO$_2$ crystal, the crystal temperature increases proportionally by $\Delta T \propto E / C(T)$. This temperature rise is measured as the change in voltage across the biased NTD thermistor. The resulting detector response is a thermal pulse with a characteristic shape that is determined by the thermal couplings and capacitances of the calorimeter components.

\begin{figure}[h]
  \centering
  \includegraphics[trim={0cm 1.6cm 0cm 0cm},clip,width=\columnwidth]{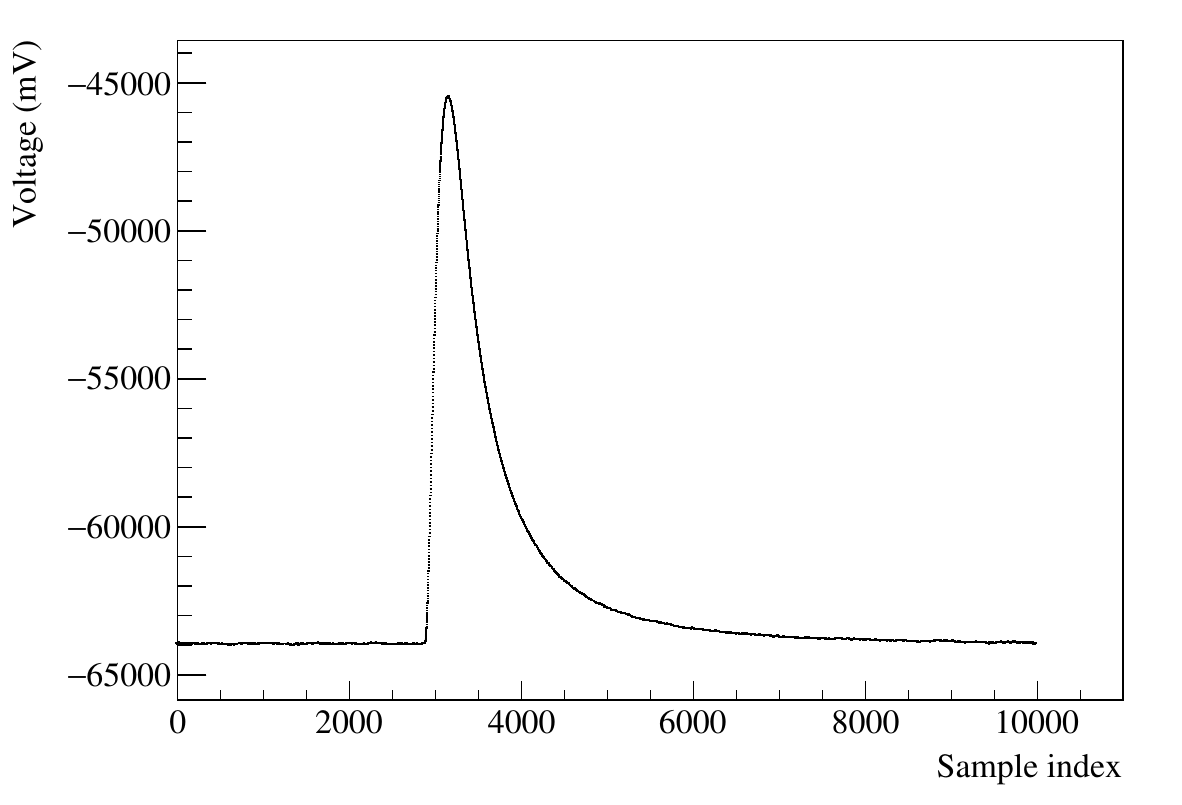}
  \includegraphics[width=\columnwidth]
  {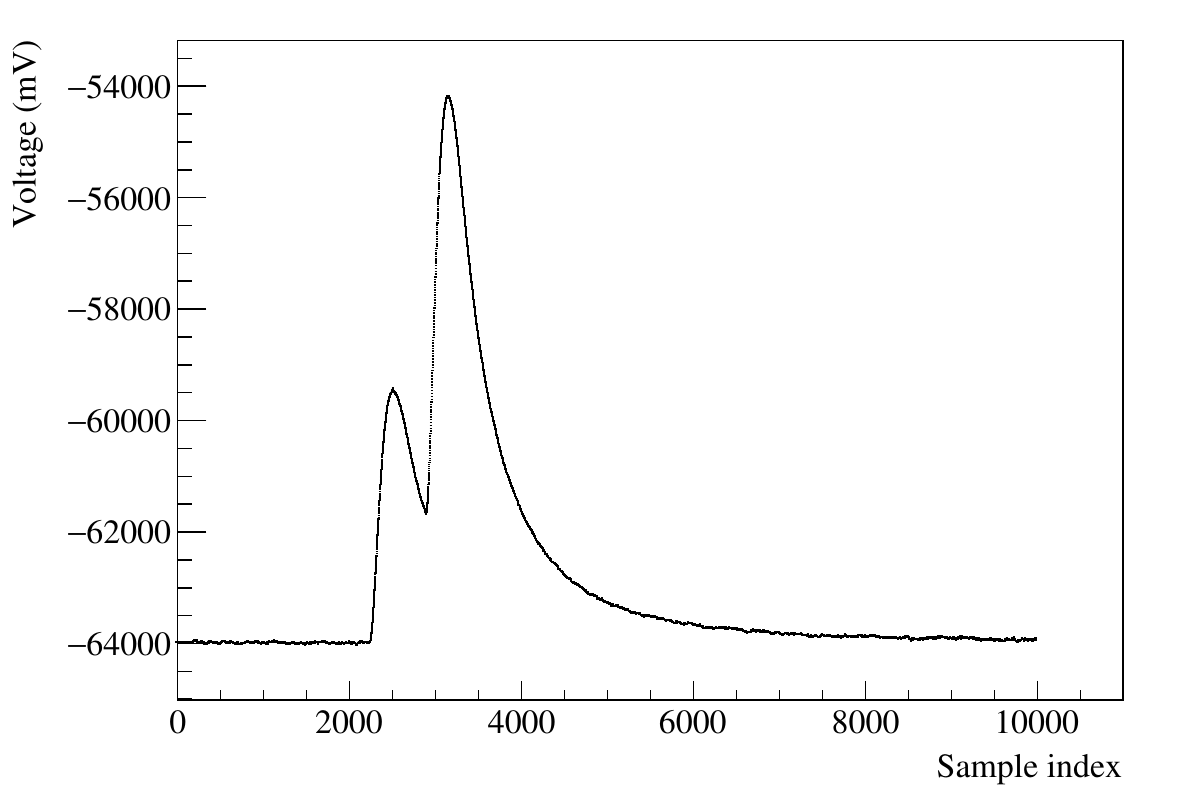}
  \caption{Representative CUORE detector pulses. \textbf{Top:} A clean single-pulse event illustrating the characteristic fast rise ($\sim$50~ms) and slow exponential
  decay ($\sim$1~s) of the CUORE cryogenic calorimeters. \textbf{Bottom:} A two-pulse pile-up event in which the triggered pulse begins while an earlier pulse is still decaying, producing a distorted double-peaked waveform. 
  }
  \label{fig:pulses}
\end{figure}

Typical CUORE pulses have a rise time of $\sim$50~ms and a thermalization (decay) time of $\sim$1~s. The continuous data stream is formed by passing the thermistor voltage through a
room-temperature preamplifier and anti-aliasing filter, and digitizing at 1~kHz. A representative clean pulse and pile-up pulse are shown in Figure~\ref{fig:pulses}.

\subsection{Pile-Up Events}

Pile-up occurs when two or more separate interactions deposit energy in a single crystal calorimeter, within an event window of 10~s. 
In CUORE, pile-up events are a source of background in the 0$\nu\beta\beta$ decay search because they can mimic the energy of a single-particle event or misrepresent the background near $Q_{\beta\beta}$, where the posited 0$\nu\beta\beta$ decay signature peak is expected.

\section{Dataset Description}

\subsection{Data Content}

The dataset consists of two classes of CUORE detector pulses:

\begin{itemize}
  \item \textbf{Class 0 -- Clean pulses:} Single-pulse events exhibiting the standard CUORE thermal pulse shape within the 10-s event window. 
  This type of event typically corresponds to an energy deposition from a single particle.

  \item \textbf{Class 1 -- Pile-up pulses:} Events in which two or more pulses are present within the 10-s event window. The pulses are time-separated and may have different amplitudes.
\end{itemize}

Each data point contains the fields listed in Table~\ref{tab:fields}. The pulse array is the raw thermistor voltage, sampled at 1~kHz, as a function of time. Normalization parameters associated with each pulse are included to expedite standard AI/ML data preprocessing.

\begin{table*}[t]
\caption{Description of fields contained in each data point of this release.}
\label{tab:fields}
\begin{ruledtabular}
\begin{tabular}{llll}
\textbf{Field} & \textbf{Description} & \textbf{Data Type} & \textbf{Notes} \\
\hline
\texttt{pulse}          & Detector pulse waveform         & \texttt{array(size=10000, dtype=float32)} & Time-series at 1~kHz \\
\texttt{label}          & Classification label            & \texttt{int8}   & 0 = clean, 1 = pile-up \\
\texttt{normalizationOffset} & 
Pulse baseline      & \texttt{float32} & In ADC units \\
\texttt{normalizationMaximum} & 
Maximum value in event window  & \texttt{float32} & In ADC units \\
\texttt{normalizationScale}  & 
Scale factor for unity max. value & \texttt{float32} & In ADC units \\
\texttt{id}             & Unique 
identifier         & \texttt{int32}  & Global, zero-indexed \\
\end{tabular}
\end{ruledtabular}
\end{table*}

\subsection{Pulse Array Format}

Each pulse is stored as a one-dimensional NumPy array of length 10000, corresponding to a 10-s event window. The window is defined such that the triggered pulse rise midpoint occurs at approximately index 3000, providing $\sim$3~s of pre-trigger baseline~\cite{CUOREtech}. The full window captures both the rising edge and a sufficient portion of the decaying tail for pulse shape characterization. Notably, each triggered pulse is associated with its own event window, so a set of pile-up pulses may be present in multiple event windows.

\subsection{Normalization Parameters}

Each pulse in the dataset is provided with a pair of normalization parameters. For each event $i$ with stored raw waveform $p_i(t)$, a corresponding normalized waveform can be constructed as

\begin{equation}
  \tilde{p}_i(t) = \frac{p_i(t) - \mu_i}{s_i + \varepsilon},
  \label{eq:norm}
\end{equation}

\noindent where $\mu_i$ is the \texttt{normalizationOffset}, $s_i$ is the \texttt{normalizationScale}, and $\varepsilon = 10^{-8}$ is a small constant to prevent division by zero. 
\noindent 
The per-event baseline ($\mu_i$) is estimated as the mean of the first $N_\text{base}$ samples of the raw waveform:

\begin{equation}
  \mu_i = \frac{1}{N_\text{base}}
    \sum_{t=0}^{N_\text{base}-1} p_i(t),
\end{equation}

\noindent where $N_\text{base} = 2500$. 
The normalization scale factor, $s_i$ is
\begin{equation}
  s_i = m_i - \mu_i
\end{equation}
where $m_i$ is the \texttt{normalizationMaximum}, the maximum raw waveform value in the event window. These three normalization parameters are stored alongside each event in the HDF5 file (see Table~\ref{tab:fields}).

\subsection*{Labels}

The binary classification label $\ell_i$, is derived from the number of flagged pulses ($n_{\text{pulses},i}$) in the event window and the CUORE 2~tonne$\cdot$yr pulse shape discrimination cut which uses the event-based Normalized Reconstruction Error~($nre_i$) to identify pulses whose shapes are compatible with the expected particle-induced thermal pulse~\cite{CUOREtech}: 

\begin{equation}
  \ell_i =
  \begin{cases}
    0, & \text{if } n_{\text{pulses},i} = 1\ \&\&\ |nre_i| < 10 \quad \text{(clean pulse)}; \\
    1, & \text{if } n_{\text{pulses},i} > 1\ \&\&\ |nre_i| > 10 \quad \text{(pile-up)}.
  \end{cases}
\end{equation}\label{eq:label}

\begin{table}[h]
\caption{Summary of the CUORE AI/ML data release.}
\label{tab:summary}
\begin{ruledtabular}
\begin{tabular}{lrr}
\textbf{Property} & \textbf{Value} \\
\hline
Total data points (Events)        & \texttt{10000} \\
Clean pulse events (Class 0)   & \texttt{5000} \\
Pile-up events (Class 1) & \texttt{5000} \\
Waveform length (Samples) & \texttt{10000} \\
Sampling rate            & 1~kHz \\
Training set             & \texttt{90}\% \\
Test set                 & \texttt{10}\% \\
File format              & HDF5 (.h5) \\
DOI                      & \texttt{10.5281/zenodo.20721645~\cite{zen}} \\
\end{tabular}
\end{ruledtabular}
\end{table}

\section{Dataset Access}

The dataset is stored in HDF5 format (\texttt{.h5}) and is accessible via Zenodo at \url{https://doi.org/10.5281/zenodo.20721645}~\cite{zen}. HDF5 files can be read in Python using the \texttt{h5py} library~\cite{h5py}.

A minimal Python example for loading and preprocessing the data is provided below. The script includes instructions for applying the normalization:

\begin{lstlisting}[
  language=Python,
  caption={Loading and normalizing the CUORE training dataset},
  label={list:loadCuoreTraining}
]
import h5py
import numpy as np

with h5py.File("cuoreTraining.h5", "r") as f:
    # Raw, unnormalized waveforms: shape (N, T)
    wf_raw = f["Waveform"][:]

    # Classification labels: shape (N,)
    # 0 = clean pulse, 1 = pile-up pulse
    labels = f["Labels"][:]

    # Per-waveform normalization parameters: shape (N, 1)
    offset = f["normalizationOffset"][:]
    maximum = f["normalizationMaximum"][:]
    scale = f["normalizationScale"][:]

# Definitions:
# offset  = mean of the first 2500 samples
# maximum = maximum raw waveform value
# scale   = maximum - offset

epsilon = 1e-8

# Construct normalized waveforms
wf_norm = (wf_raw - offset) / (scale + epsilon)

# Enforce a strict range of [0, 1]
wf_norm = np.clip(wf_norm, 0.0, 1.0)

# Add a channel dimension for Conv1D input
# New shape: (N, T, 1)
wf_norm = wf_norm[..., np.newaxis]
\end{lstlisting}

\vspace{\baselineskip}
\vspace{\baselineskip}
\vspace{\baselineskip}
\vspace{\baselineskip}
\vspace{\baselineskip}
\vspace{\baselineskip}
\vspace{\baselineskip}
\begin{lstlisting}[
  language=Python,
  caption={Loading and normalizing the CUORE test dataset},
  label={list:loadCuoreTest}
]
import h5py
import numpy as np

with h5py.File("cuoreTest.h5", "r") as f:
    # Raw, unnormalized waveforms: shape (N, T)
    wf_raw = f["Waveform"][:]

    # Per-waveform normalization parameters: shape (N, 1)
    offset = f["normalizationOffset"][:]
    maximum = f["normalizationMaximum"][:]
    scale = f["normalizationScale"][:]

# The test dataset intentionally contains no labels.

# Definitions:
# offset  = mean of the first 2500 samples
# maximum = maximum raw waveform value
# scale   = maximum - offset

epsilon = 1e-8

# Construct normalized waveforms
wf_norm = (wf_raw - offset) / (scale + epsilon)

# Enforce a strict range of [0, 1]
wf_norm = np.clip(wf_norm, 0.0, 1.0)

# Add a channel dimension for Conv1D input
# New shape: (N, T, 1)
wf_norm = wf_norm[..., np.newaxis]
\end{lstlisting}

\section{AI/ML Challenge}

To encourage engagement with this dataset, we propose a pulse classification challenge
with the following task:\\

\noindent \textbf{Classification Task:} Develop a machine learning model that takes a raw pulse
array as input and predicts the binary label (0 = clean, 1 = pile-up). Models will be
evaluated on the held-out test set using the following metrics:

\begin{itemize}
  \item \textbf{AUC-ROC}: Area under the receiver operating characteristic curve.
  \item \textbf{Accuracy}: Fraction of correctly classified events.
  \item \textbf{False Positive Rate at fixed True Positive Rate}: Specifically, the
  false positive rate at a true positive rate of 90\%, relevant for physics analyses where signal efficiency must be controlled.
\end{itemize}

Participants are invited to submit predictions on the test set (where labels are withheld) to vis224@pitt.edu. The submission format should be a CSV file or NumPy array
with one predicted probability per test event (probability of Class 1 / pile-up).

\section{Disclaimer}

The CUORE Collaboration has authorized the public release of this dataset, permitting its use for all reasonable purposes. Individuals or groups are permitted to publish results based on this dataset. The CUORE Collaboration maintains ownership rights over this dataset and reserves all associated rights. Users of this dataset are kindly requested to cite Ref.~\cite{CUORE2tyr} and this arXiv paper in any resulting
publications.

\section{Acknowledgments}
The CUORE Collaboration thanks the directors and staff of the Laboratori Nazionali del Gran Sasso and the technical staff of our laboratories. This work was supported by the Istituto Nazionale di Fisica Nucleare (INFN); the National Science Foundation under Grant Nos. NSF-PHY-0605119, NSF-PHY-0500337, NSF-PHY-0855314, NSF-PHY-0902171, NSF-PHY-0969852, NSF-PHY-1307204, NSF-PHY-1314881, NSF-PHY-1401832, NSF-PHY-1913374, and NSF-PHY-2412377; Yale University, Johns Hopkins University, and University of Pittsburgh.
This material is also based upon work supported by the US Department of Energy (DOE) Office of Science under Contract Nos. DE-AC02-05CH11231, and DE-AC52-07NA27344; by the DOE Office of Science, Office of Nuclear Physics under Contract Nos. DE-FG02-08ER41551, DE-FG03-00ER41138, DE-SC0012654, DE-SC0020423, DE-SC0019316, and DE-SC0011091.
This research used resources of the National Energy Research Scientific Computing Center (NERSC).
This work makes use of both the DIANA data analysis and APOLLO data acquisition software packages, which were developed by the CUORICINO, CUORE, LUCIFER, and CUPID-0 Collaborations.
The authors acknowledge the Advanced Research Computing at Virginia Tech and the Yale Center for Research Computing for providing computational resources and technical support that have contributed to the results reported within this paper.\\

\appendix
\section{Technical Details of the Data Release}

Because the pulse shapes can change for different detector operating conditions and are largely dependent on the individual calorimeter components, the 10000 CUORE waveforms provided in this data release come from a single CUORE dataset-calorimeter. Only the calibration runs of the single CUORE dataset were used in the event selection. Using calibration runs ensured adequate statistics of pile-up pulse events since radioactive sources are deployed around the detector during these measurements. For the event selection, basic quality cuts were applied to exclude baseline instabilities and only triggered pulses with reconstructed energy $>$500~keV were accepted.



\begin{thebibliography}{99}

\bibitem{Majorana1937}
E.~Majorana, \textit{Nuovo Cim.} \textbf{14}, 171 (1937).

\bibitem{Sakharov1967}
A.~D.~Sakharov, \textit{JETP Lett.} \textbf{5}, 24 (1967).

\bibitem{CUORE2022Nature}
D.~Q.~Adams \textit{et al.} (CUORE Collaboration),
\textit{Nature} \textbf{604}, 53 (2022).

\bibitem{CUORE2tyr}
D.~Q.~Adams \textit{et al.} (CUORE Collaboration),
\textit{Science} \textbf{390}, 1029-1032 (2025).

\bibitem{NAIRR2023}
National Science Foundation and OSTP,
``Strengthening and Democratizing the U.S. AI Innovation Ecosystem'' (2023),
\url{https://www.ai.gov/wp-content/uploads/2023/01/NAIRR-TF-Final-Report-2023.pdf}.

\bibitem{fiorini}
E.~Fiorini and T.~O.~Niinikoski, \textit{Nucl. Instrum. Meth. A} \textbf{224}, 83 (1984).


\bibitem{Haller1984}
E.~E.~Haller \textit{et al.},
in \textit{Neutron Transmutation Doping of Semiconductor Materials},
p.~21, Springer, 1984.

\bibitem{CUOREtech}
CUORE Collaboration, ``End-to-End Data Analysis Methods for the CUORE Experiment,'' arXiv:2510.25720 (2025).

\bibitem{zen}
V.~Sharma and K.~Alfonso, ``CUORE Data Release for ML Applications: Pulse Shape Analysis Dataset,'' Zenodo (2026); \url{https://doi.org/10.5281/zenodo.20721645}.

\bibitem{h5py}
A.~Collette, \textit{Python and HDF5}, O'Reilly, 2013.
\url{https://www.h5py.org}.

\end{thebibliography}
\end{document}